\documentclass{article}
\usepackage{spconf,amsmath,epsfig, amssymb, graphicx, dsfont, setspace,url, amsfonts, amsthm}

\title{CAUSAL COMPRESSIVE SENSING FOR GENE NETWORK INFERENCE}
\name{Mo Deng, Amin Emad, and Olgica Milenkovic
\thanks{This work was supported by the NSF STC-CSoI 2011 and NSF CCF 0809895 grants and a AFRLDL-EBS AFOSR Complex Networks grant.}}
\address{Department of Electrical and Computer Engineering \\
University of Illinois, Urbana-Champaign\\ e-mail:  \{modeng1,emad2,milenkov\}@illinois.edu
}
\begin{document}
%
\maketitle
\begin{abstract}
We propose a novel framework for studying causal inference of gene interactions using
a combination of compressive sensing and Granger causality techniques. The gist
of the approach is to discover sparse linear dependencies between time series of gene expressions via a Granger-type
elimination method. The method is tested on the Gardner dataset for the SOS network in \emph{E. coli}, for which both known and unknown causal 
relationships are discovered.
\end{abstract}
\vspace{-0.08in}
\section{Introduction}\label{sec:intro}
One of the focal problems of systems biology is the discovery of causal
relationships among different components of biological systems. Gene
regulatory networks, protein-protein interaction networks, chemical signaling, and metabolic networks all exhibit 
causal relationships between their agents that are crucial for proper functionality. Discovering such causal relationships through experiments may be a challenging 
task due to the technical precision required from the experiments and due to the large number of interconnected and dynamically
varying components of the system. It is therefore of great importance to develop an analytical framework for discovering causal
connections between genes and for elucidating the gene interactome, based on limited experimental data. 
Analytically inferred interactions may consequently be used to guide the experimental design process, which would
then help in further refining the modeling framework. Representative work in this direction includes~\cite{SM10,RHSE08}.

One way to detect if a gene causally influences another gene is to observe the target gene's expression
levels and detect if changes in the expressions of the other gene affect changes in the expressions of the target. 
For this purpose, a number of authors suggested the use of Bayesian
network modeling and other machine learning tools, algebraic techniques, information-theoretic
methods, and autoregressive Granger causality~\cite{G69} approaches.

We propose a novel method for identifying causal gene dependencies based on two ideas: compressive sensing and a non-standard version of
Granger causality, or elimination analysis. The compressive sensing approach is motivated by a technique for face recognition used in computer vision, first described in~\cite{WMMSHY10}.
The crux of the approach is to efficiently find a sparse linear representation of an image of one individual in terms of images of other individuals and the individual itself,
taken under many different conditions. This setup is reminiscent to the one described in~\cite{HDW09,PFLL11}, where expression levels of genes taken under different experimental 
conditions (or, under different gene knockout scenarios) are represented as vectors for which a sparse representation is sought.

The main finding of our analysis indicates that even without additional learning steps, one can infer a number of causal gene interactions in the \emph{E. coli} SOS network
reported by Gardner et al.~\cite{GBLC03} using a combination of elimination techniques and
compressive sensing, even when the number of points in the time series is very small. Causality is inferred via a reduction in the representation residual error and the presence of a certain
gene as a factor in a sparse representation. Granger causality may be incorporated in this framework
by using time delayed profiles for recognition purposes and incorporating them into the sensing matrix as part of an elimination approach. 
Unfortunately, the proposed method may not yield to improved detection probability upon adding time shifted expression profiles, which may be attributed to 
the fact that gene expressions are usually measured at time instances that are too widely separated.

The paper is organized as follows. Compressive sensing and Granger causality are briefly introduced in Section~\ref{sec:background}. The proposed causal inference method 
is described in Section~\ref{sec:model}, while the testing framework and the results pertaining to the gene SOS network of \emph{E. coli} are presented in Section~\ref{sec:results}.

\vspace{-0.08in}
\section{BACKGROUND}
\label{sec:background}
\vspace{-0.08in}

Compressive sensing (CS) is a technique for efficient sampling of compressible and $K$-sparse signals,
i.e., signals that can be represented by $K\ll N$ significant coefficients
over a $N$-dimensional basis. Sampling of a $K$-sparse, discrete-time
signal $\mathbf{x}$ of dimension $N$ is accomplished by computing
a measurement vector $\textbf{y}$ that consists of $m\ll N$ linear
projections, i.e., 
\begin{equation}\label{eq:sensing}
\mathbf{y}=\mathbf{\Phi}\mathbf{x}.
\end{equation}
Here, $\mathbf{\Phi}$ represents an $m\times N$ matrix, usually
over the field of real numbers with certain desirable singular value properties~\cite{CW08,D06}. One of these properties is known
as the restricted isometry property (RIP), and it is usually considered a benchmark condition for practical testing of compressive sensing
methods. Furthermore, the recent results in  \cite{CW08} and \cite{D06} 
showed that a length $N$ signal that is $K$-sparse can be recovered from only $m\sim O(K\log{(N/K)})$ linear observations of the signal,
provided that $\boldsymbol{\Phi}$ has the RIP.

Only a few compressive sensing algorithms were successfully integrated into causal inference models~\cite{SM10}. One of the 
simplest causality testing schemes, originally proposed in econometrics, is Granger causality. In its original incarnation, Granger causality was presented as 
a heuristic statistical concept based on prediction. The method has the goal to determine if a time series of past observations of a process helps to predict the future values of another process. 
More formally, a process $X$ Granger-causes another process $Y$ if the future values of $Y$ may be predicted with larger accuracy using the past observations of both $Y$ and $X$ 
than when using only past observations of $Y$. In the context of linear regression, this causality model may be described as follows. Let us assume that the value of a process 
$Y$ at time $t$ may be predicted via an autoregressive (linear) rule as
\begin{align} \label{eq:granger}
Y^{(1)}(t)=\sum_{i=1}^{d}a_{i}Y(t-iT)+r^{(1)}, 
\end{align}
where the prediction memory is denoted by $d$ and the sampling interval by $T$. The residual error of the predictor equals $r^{(1)}$. Now, let
\begin{align}
Y^{(2)}(t)=\sum_{i=1}^{d}a'_{i}Y(t-iT)+\sum_{i=1}^{d}b_{i}X(t-iT)+r^{(2)}
\end{align}
be a linear prediction of $Y$ at time $t$ when the past values of $X$ are considered in addition to the observation of $Y$. The residual error in this case equals $r^{(2)}$.
One way to determine if $X$ Granger-causes $Y$ is to compare the residual errors: if the value of $r^{(2)}$ is smaller than the value of $r^{(1)}$, then $X$ Granger-causes $Y$. 
The word ``smaller"' is usually interpreted in many different ways, frequently involving constraints other than just the difference of the residuals.

In what follows, we propose to combine the techniques captured by \eqref{eq:sensing} and \eqref{eq:granger} for the purpose of inferring causal relationships in gene regulatory networks. There are
two main issues to be addressed in this case: how to discover linear relationships between expression profiles that may (and usually are) correlated with each other and how to adapt the sensing
matrix $\boldsymbol{\Phi}$ to perform meaningful Granger-type tests. We discuss these two issues in what follows.

\vspace{-0.08in} 
\section{THE MODEL} \label{sec:model}

We start by briefly describing the work~\cite{WMMSHY10}  that illustrates how compressive sensing may be used for pattern recognition in the presence of noise and outliers. Assume that one wants to identify a 
person, given a particular image of the person, using a large database of images of different individuals taken under different conditions. One way to perform the identification is to convert each
image into a vector, the target vector being $\mathbf{y}$, and the database vectors being $\boldsymbol{\phi}_i$, $i=1,\ldots,N$, where $N$ denotes the number of individuals in the database.
The sensing matrix equals $\mathbf{\Phi}=[\boldsymbol{\phi}_1,\boldsymbol{\phi}_2,\cdots,\boldsymbol{\phi}_{N}]$ and recognition is performed by solving \eqref{eq:sensing} for a judiciously 
chosen sparsity level $K$ of the vector $\mathbf{x}$. Then, one can decide on the identity of
the target individual based on how many of the $K$ non-zero components of $\mathbf{x}$ correspond to images of the target individual.

Consider the following related scenario: instead of dealing with images, we focus on expression profiles of $N$ genes taken under different experimental conditions. These expression
time series now represent the columns of the sensing matrix $\boldsymbol{\phi}_{g_i}$, $i=1,\ldots,N$. Due to the fact that the task at hand in gene network analysis is not to identify a gene based
on its expression, but rather genes that influence its behavior or are co-expressed, the setup has to be changed slightly. In this case, one should declare one gene of interest to be the 
\textbf{target gene}, with expression profile $\mathbf{y}$, and the goal would be to identify correlated genes via \eqref{eq:sensing}. A similar approach to this one was pursued in~\cite{PFLL11},
with the goal of identifying linear dependencies between expression vectors based on sparse interaction assumptions. Compressive sensing was used only as an initial step of a learning
procedure, implemented via the AdaBoost method.

In order to incorporate Granger causality into the above described framework, let us assume that $G$ denotes the set of all possibly interacting genes in a transcription network. 
We perform two compressive sensing tests. First, we find the ``relevant'' genes for target gene $j$ when the expression profile of gene $i$ is not present in the sensing matrix. In this case
\begin{align}
\mathbf{y}_j^{(1)}=\Phi_{G\backslash \{i,j\}}\mathbf{x}_j^{(1)},
\end{align}
where $\mathbf{x}_j^{(1)}$ denotes the sparse vector of coefficients, and $\Phi_{G\backslash \{i,j\}}$ is an $m\times n$ matrix representing the gene expressions of all the genes in $G$ except genes $i$ and $j$, i.e.
\begin{align}
\Phi_{\!G \backslash\{i,j\}}\!=\![\boldsymbol{\phi}_{\!g_1}\!,\!\boldsymbol{\phi}_{\!g_2}\!,\!\cdots,\boldsymbol{\phi}_{\!g_{j\!-\!1}}\!,\!\boldsymbol{\phi}_{\!g_{j\!+\!1}}\!,\!\cdots\!,\!\boldsymbol{\phi}_{\!g_{i\!-\!1}}\!,\!\boldsymbol{\phi}_{\!g_{i\!+\!1}}\!,\!\cdots,\boldsymbol{\phi}_{\!g_{n}}]. \notag
\end{align} 
Second, we perform one more round of testing for gene $j$ when gene $i$ is \textbf{included in the sensing matrix},
\begin{align} \label{eq:no-exclusion}
\mathbf{y}_j^{(2)}=\Phi_{G \backslash \{j\}}\mathbf{x}_j^{(2)},
\end{align}
where 
\begin{align}
\Phi_{\!G \backslash\{j\}}\!=\![\boldsymbol{\phi}_{\!g_1}\!,\!\boldsymbol{\phi}_{\!g_2}\!,\!\cdots,
\boldsymbol{\phi}_{\!g_{j\!-\!1}}\!,\!\boldsymbol{\phi}_{\!g_{j\!+\!1}}\!,\!\cdots\!,\!\boldsymbol{\phi}_{\!g_{i\!-\!1}}\!,\!\boldsymbol{\phi}_{\!g_{i}},\!\boldsymbol{\phi}_{\!g_{i\!+\!1}}\!,\!\cdots,\boldsymbol{\phi}_{\!g_{n}}]. \notag
\end{align} 
The results of the two experiments may be used for inference as follows. If the residual error of recovering $\mathbf{y}_j^{(2)}$ is smaller than the residual error of recovering $\mathbf{y}_j^{(1)}$, and in addition,
gene $g_i$ was included in the list of $K$ non-zero components of the approximation, we conclude that $g_i$ causally influences $g_j$. We would like to point out that this is one of the simplest ways to
deduce causal relationships via the proposed method -- for more precise predictions, one may also record the number of perturbations in the corresponding $\mathbf{x}$ vector caused by including gene $g_i$, the
magnitude of the weight of gene $g_i$, etc.  A full description of these criteria is postponed to be described in the full version of the paper.

This approach may be taken one step further. Assume that $S^{i}$ denotes a right-shift operator by $i$ locations, which when applied to a vector $(x_{\ell},x_{\ell+1},\ldots,x_N)$ produces 
\[S^{i}(x_{\ell},x_{\ell+1},\ldots,x_N)=(x_{\ell-i},x_{\ell},\ldots,x_{N-i}).\]
If the sensing matrix $\boldsymbol{\Phi}$ in the latter of the two above described scenarios is chosen as
\begin{align}
\![\boldsymbol{\phi}_{\!g_1}\!,   S^{1} \boldsymbol{\phi}_{\!g_1}, \ldots,  S^{f} \boldsymbol{\phi}_{\!g_1}, \ldots, 
\boldsymbol{\phi}_{\!g_N}\!,   S^{1} \boldsymbol{\phi}_{\!g_N}, \ldots,  S^{f} \boldsymbol{\phi}_{\!g_N}], \notag
\end{align} 
for some integer $f>1$, then the compressive sensing model becomes a sparse instant of Granger causality. Although we tested this model extensively, the corresponding findings are not included
in the manuscript. This is due to the fact that including shifts of profiles does not seem to offer practical performance improvements for the proposed inference method, which we attribute to the fact that
the size of the sensing matrix $\boldsymbol{\Phi}$ has to be doubled and that the time point measurements are taken at instances too apart from each other. The latter phenomena clearly does not
allow for inferring short-range dependencies, usually found in gene regulatory networks.

Many other extensions of this modeling framework are possible - one being a model where the columns of the sensing matrix represent non-linear functions of pairs of expression vectors.
This and other models will be discussed in a companion paper.
\vspace{-0.1in}
\section{RESULTS}
\label{sec:results}

In order to test the causal compressive sensing method outlined in Section 3, we considered expression levels of $4292$ genes of the bacteria \emph{E. coli} obtained from the KEGG database, 
after removing genes with very few accurate expression points. The expression vectors correspond to $22$ experiments with $94$ time points for each gene. 
Consequently, the expression profiles may be organized into a $94 \times 4292$ real-valued matrix. Each entry in the matrix is normalized as follows: the average expression level of a column is
computed and then subtracted from each element in the column. Hence, the normalized matrix contains both positive and negative entries.
\begin{table}[t!] 
	\centering
	\caption{A subnetwork of the SOS network of \emph{E. coli}: an entry $1$ at location $(i,j)$ indicates that gene $j$ causally regulates gene $i$; $0$ indicates that no such regulation is currently known.}
	\vspace{0.07in}
		\begin{tabular}{|c||c|c|c|c|c|c|c|c|c|}
			\hline 
			 &  $\!\!$dinI $\!\!\!\!$& $\!\!\!$lexA$\!\!\!$  & $\!\!\!$recA$\!\!\!$ & $\!\!\!$recF$\!\!\!$ & $\!\!\!$rpoD$\!\!\!$ & $\!\!\!$rpoH$\!\!\!$ & $\!\!\!$rpoS$\!\!\!$ & $\!\!\!$ssb$\!\!\!$ & $\!\!\!$umuCD$\!\!\!$\\ 
			\hline\hline
			$\!\!$dinI$\!\!$  & 0 & 1& 0& 0& 1& 0& 0& 0& 0      \\
			\hline
			$\!\!$lexA$\!\!$ & 0 & 1& 0& 0& 1& 0& 0& 0& 0      \\
                            \hline
			$\!\!$recA$\!\!$ & 0 & 1& 0& 0& 1& 0& 0& 0& 0      \\
                            \hline
			$\!\!$recF$\!\!$ & 0 & 0& 0& 0& 1& 0& 1& 0& 0      \\			
                            \hline
			$\!\!$rpoD$\!\!$ & 0 & 1& 0& 0& 1& 1& 0& 0& 0      \\			
                            \hline
			$\!\!$rpoH$\!\!$ & 0 & 0& 0& 0& 1& 1& 0& 0& 0      \\					
                            \hline
			$\!\!$rpoS$\!\!$ & 0 & 0& 0& 0& 1& 0& 1& 0& 0      \\	
                            \hline
			$\!\!$ssb$\!\!$ & 0 & 1& 0& 0& 1& 0& 0& 0& 0      \\	
                            \hline
			$\!\!$umuCD$\!\!$ & 0 & 1& 0& 0& 1& 0& 0& 0& 0      \\										
						\hline					

			\end{tabular}\label{table:Gardner}
\end{table}

The test network of interest was chosen to be the network induced by genes of the SOS repair system. Although this system has at least fifty documented components, we focused on nine genes deemed 
most relevant by the analysis in~\cite{GBLC03} (see Table I for the description of the topology of this network). One of the key genes in this network, named \emph{lexA}, is known to regulate many genes in the SOS system and was chosen as the starting target gene of our analysis. 

Using the setup of eq.~\eqref{eq:no-exclusion}, one in which no gene is excluded, we find $K=50$ relevant genes for \emph{lexA}. 
Since a few genes known to be present in the SOS network were not in the identified list, we added them back into the pool. The reason for performing this step is to reduce the search space and to
clean up the potential list of candidates for interaction partners of \emph{lexA}. A very large number of genes included in the analysis increases the interference level and sensitivity of the procedure.
To verify that this expurgation procedure is meaningful, we computed the $p$-value of SOS gene inclusion, which equals $0.0036$. At the end of the procedure, we were left with $57$ genes.

 \begin{table}[t!]
	\centering
	\caption{Table of residues obtained via Granger testing, reported as (\emph{res2}-\emph{res1})/\emph{res1} ($\%$). The largest changes in the residues detected are listed in boldface script.}
	\vspace{0.07in}
		\begin{tabular}{|c||c|c|c|c|c|c|c|c|c|}
			\hline 
			 &  $\!\!$dinI $\!\!\!\!$& $\!\!\!$lexA$\!\!\!$  & $\!\!\!$recA$\!\!\!$ & $\!\!\!$recF$\!\!\!$ & $\!\!\!$rpoD$\!\!\!$ & $\!\!\!$rpoH$\!\!\!$ & $\!\!\!$rpoS$\!\!\!$ & $\!\!\!$ssb$\!\!\!$ & $\!\!\!$umuCD$\!\!\!$\\ 
			\hline\hline
			$\!\!$dinI$\!\!$  & $\!\!$\footnotesize{0}$\!\!$ & $\!\!$\footnotesize{-1}$\!\!$& $\!\!$\footnotesize{1}$\!\!$& $\!\!$\footnotesize{4}$\!\!$& $\!\!$\footnotesize{14.5}$\!\!$& $\!\!$\footnotesize{6.6}$\!\!$& $\!\!$\footnotesize{2.3}$\!\!$& $\!\!$\footnotesize{9.2}$\!\!$& $\!\!$\footnotesize{5.1}$\!\!$      \\
			\hline
			$\!\!$lexA$\!\!$ & $\!\! $\footnotesize{\textbf{-14.3}}$ \!\!$ & \footnotesize{0}& \footnotesize{0}& $\!\!$\footnotesize{0.4}$\!\!$& \footnotesize{0}& \footnotesize{0}& \footnotesize{0}& $\!\!$\footnotesize{1.5}$\!\!$& $\!\!$\footnotesize{-3.6}$\!\!$      \\
                            \hline
			$\!\!$recA$\!\!$ & $\!\!$\footnotesize{\textbf{-13}}$\!\!$ & $\!\!$\footnotesize{0.2}$\!\!$& \footnotesize{0}& $\!\!$\footnotesize{5.2}$\!\!$& $\!\!$\footnotesize{1}$\!\!$& $\!\!$\footnotesize{3.3}$\!\!$& $\!\!$\footnotesize{\textbf{-7.1}}$\!\!$& \footnotesize{0}& $\!\!$\footnotesize{-1.7}$\!\!$      \\
                            \hline
			$\!\!$recF$\!\!$ & $\!\!$\footnotesize{-3.6}$\!\!$ & $\!\!$\footnotesize{5.5}$\!\!$& $\!\!$\footnotesize{-6.3}$\!\!$& \footnotesize{0}& $\!\!$\footnotesize{\textbf{-20}}$\!\!$& $\!\!$\footnotesize{2.4}$\!\!$& $\!\!$\footnotesize{-2}$\!\!$& \footnotesize{0}& \footnotesize{0}      \\			
                            \hline
			$\!\!$rpoD$\!\!$ & $\!\!$\footnotesize{-2.6}$\!\!$ & $\!\!$\footnotesize{-4.7}$\!\!$& $\!\!$\footnotesize{-2.3}$\!\!$& $\!\!$\footnotesize{\textbf{-19}}$\!\!$& $\!\!$\footnotesize{0}$\!\!$& $\!\!$\footnotesize{\textbf{-9}}$\!\!$& $\!\!$\footnotesize{-2.4}$\!\!$& $\!\!$\footnotesize{\textbf{-8.9}}$\!\!$& $\!\!$\footnotesize{-2.4}$\!\!$  \\			
                            \hline
			$\!\!$rpoH$\!\!$ & $\!\!$\footnotesize{4}$\!\!$ & $\!\!$\footnotesize{0.1}$\!\!$& $\!\!$\footnotesize{10}$\!\!$& $\!\!$\footnotesize{-4}$\!\!$& $\!\!$\footnotesize{\textbf{-9}}$\!\!$& $\!\!$\footnotesize{0}$\!\!$& $\!\!$\footnotesize{19}$\!\!$& $\!\!$\footnotesize{24}$\!\!$&$\!\!$\footnotesize{-0.7}$\!\!$      \\					
                            \hline
			$\!\!$rpoS$\!\!$ & $\!\!$\footnotesize{\textbf{-22}}$\!\!$ & $\!\!$\footnotesize{20}$\!\!$& $\!\!$\footnotesize{11}$\!\!$& $\!\!$\footnotesize{4}$\!\!$& $\!\!$\footnotesize{4}$\!\!$& $\!\!$\footnotesize{3}$\!\!$& $\!\!$\footnotesize{0}$\!\!$& $\!\!$\footnotesize{0.1}$\!\!$& $\!\!$\footnotesize{21}$\!\!$      \\	
                            \hline
			$\!\!$ssb$\!\!$ & $\!\!$\footnotesize{-0.5}$\!\!$ & $\!\!$\footnotesize{-3}$\!\!$& $\!\!$\footnotesize{-0.3}$\!\!$& $\!\!$\footnotesize{11}$\!\!$& $\!\!$\footnotesize{11}$\!\!$& $\!\!$\footnotesize{8}$\!\!$& $\!\!$\footnotesize{13}$\!\!$& $\!\!$\footnotesize{0}$\!\!$& $\!\!$\footnotesize{-0.3}$\!\!$      \\	
                            \hline
			$\!\!$umuCD$\!\!$ & $\!\!$\footnotesize{-1.1}$\!\!$ & $\!\!$\footnotesize{-0.3}$\!\!$& $\!\!$\footnotesize{0.5}$\!\!$& $\!\!$\footnotesize{0}$\!\!$& $\!\!$\footnotesize{6}$\!\!$& $\!\!$\footnotesize{0}$\!\!$& $\!\!$\footnotesize{0}$\!\!$& $\!\!$\footnotesize{0}$\!\!$& $\!\!$\footnotesize{0}$\!\!$      \\										
						\hline					

			\end{tabular}\label{table:Residue}
\end{table}

We then proceeded to the second stage of identification. We used eq.~\eqref{eq:no-exclusion} for a second time, with $K=5$, but applied the procedure to all $57$ genes identified in the first step as target genes. 
We formed a union of  genes relevant for each SOS network gene (each set of five relevant genes included at least two SOS network genes) and each out-of-network gene. This reduced the test
set to 34 test genes. These genes were then used in the elimination method described in the previous section, i.e., for each of the $34$ test genes a Granger
test was performed with respect to the remaining $33$ genes. The residuals of the reconstruction without and with a given gene are denoted by $res1$ and $res2$, respectively, 
and which genes these residuals correspond to
should be clear from the context. 
The Granger tests are performed for a range of values $K=5$ to $K=20$. The residuals obtained for $K=14$ are listed in Table 2. 
As already pointed out, the list of relative residual changes
does not suffice to determine if a gene causally influences another gene. One also has to verify that the elimination gene was identified as relevant when included in the compressive sensing process. 
The results of this combined test are presented in Table 3.

\begin{table}[t!]
	\centering
	\caption{``U'' stands for ``unconfirmed'' and points to links in the Gardner network that were not found by our method;``P'' stands for ``predicted'' and 
	corresponds to new links found using our method that were not previously reported in Gardner's paper; ``V'' stands for ``verified'' and refers to links reported in Gardner's network and verified by our method.}
	\vspace{0.07in}
		\begin{tabular}{|c||c|c|c|c|c|c|c|c|c|}
			\hline 
			 &  $\!\!$dinI $\!\!\!\!$& $\!\!\!$lexA$\!\!\!$  & $\!\!\!$recA$\!\!\!$ & $\!\!\!$recF$\!\!\!$ & $\!\!\!$rpoD$\!\!\!$ & $\!\!\!$rpoH$\!\!\!$ & $\!\!\!$rpoS$\!\!\!$ & $\!\!\!$ssb$\!\!\!$ & $\!\!\!$umuCD$\!\!\!$\\ 
			\hline\hline
			$\!\!$dinI$\!\!$  & 0 & $\!\!$$1_V$$\!\!$& 0& 0& $\!\!$$0_U$$\!\!$& 0& 0& 0& 0      \\
			\hline
			$\!\!$lexA$\!\!$ & $\!\!$$1_P$$\!\!$ & 0 & 0& 0& $\!\!$$0_U$$\!\!$ & 0& 0& 0& $\!\!$$1_P$$\!\!$      \\
                            \hline
			$\!\!$recA$\!\!$ & $\!\!$$1_P$$\!\!$ & $\!\!$$1_V$$\!\!$& 0& 0& $\!\!$$1_V$$\!\!$& 0& 0& 0& 0      \\
                            \hline
			$\!\!$recF$\!\!$ & 0 & 0& 0& 0& $\!\!$$1_V$$\!\!$& 0& $\!\!$$0_U$$\!\!$& 0& 0      \\			
                            \hline
			$\!\!$rpoD$\!\!$ & 0 & $\!\!$$1_V$$\!\!$& 0& $\!\!$$1_P$$\!\!$& 0& $\!\!$$0_U$$\!\!$& 0& 0& $\!\!$$1_P$$\!\!$      \\			
                            \hline
			$\!\!$rpoH$\!\!$ & 0 & 0& 0& $\!\!$$1_P$$\!\!$& $\!\!$$1_V$$\!\!$& 0& 0& 0& 0      \\					
                            \hline
			$\!\!$rpoS$\!\!$ & $\!\!$$1_P$$\!\!$ & 0& 0& 0& $\!\!$$0_U$$\!\!$& 0& 0& 0& 0      \\	
                            \hline
			$\!\!$ssb$\!\!$ & $\!\!$$1_P$$\!\!$ & $\!\!$$1_V$$\!\!$& 0& 0& $\!\!$$0_U$$\!\!$& 0& 0& 0& 0      \\	
                            \hline
			$\!\!$umuCD$\!\!$ &  $\!\!$$1_P$$\!\!$ & $\!\!$$1_V$$\!\!$& 0& 0& $\!\!$$0_U$$\!\!$& 0& 0& 0& 0    \\										
						\hline					

			\end{tabular}\label{table:LinkDetection}
\end{table}

Due to space limitations, we comment only on one result presented in Table 3. The result concerns the genes \emph{dinI} and \emph{recA}. The network of Gardner, published in 2003, did not
include causal dependencies involving \emph{dinI} as a regulator of  \emph{recA}. Before 2003, it was experimentally verified that \emph{recA} leads to cleavage of a repressor of \emph{lexA}, but only
in the last few years did it become apparent that  \emph{dinI} has a strong role in regulating \emph{recA}'s ability to promote \emph{lexA} cleavage. 
Note that this dependency was detected with a very large reduction in the residual error, equal to $13\%$.

We conclude our exposition with one more interesting finding, described in Table 4. There, we listed the ``perturbation'' in the list of relevant genes caused by the inclusion of a particular gene after its initial
exclusion from analysis. The perturbation caused by the inclusion of the \emph{dinI} gene into the sensing matrix of the \emph{recA} is significant: 9 genes (out of 14).
\vspace{-0.1in}
 \begin{table}[t!]
	\centering
	\caption{The number of changes in relevant genes induced by the elimination process.}
	\vspace{0.05in}
		\begin{tabular}{|c||c|c|c|c|c|c|c|c|c|}

			\hline 

			 &  $\!\!$dinI $\!\!\!\!$& $\!\!\!$lexA$\!\!\!$  & $\!\!\!$recA$\!\!\!$ & $\!\!\!$recF$\!\!\!$ & $\!\!\!$rpoD$\!\!\!$ & $\!\!\!$rpoH$\!\!\!$ & $\!\!\!$rpoS$\!\!\!$ & $\!\!\!$ssb$\!\!\!$ & $\!\!\!$\small{umuCD}$\!\!\!$\\ 
			\hline\hline
			$\!\!$dinI$\!\!$  & 0 & 8& 7& 5& 4& 5& 7& 4& 7      \\
			\hline
			$\!\!$lexA$\!\!$ & 9 & 0& 6& 6& 5& 4& 5& 5& 4      \\
                            \hline
			$\!\!$recA$\!\!$ & 9 & 6& 0& 5& 6& 6& 7& 3& 3      \\
                            \hline
			$\!\!$recF$\!\!$ & 5 & 5& 5& 0& 7& 4& 4& 2& 2      \\			
                            \hline
			$\!\!$rpoD$\!\!$ & 7 & 5& 5& 5& 0& 5& 3& 4& 2      \\			
                            \hline
			$\!\!$rpoH$\!\!$ & 9 & 5& 8& 7& 5& 0& 6& 5& 5      \\					
                            \hline
			$\!\!$rpoS$\!\!$ & 10 & 11& 9& 4& 4& 5& 0& 4& 5      \\	
                            \hline
			$\!\!$ssb$\!\!$ & 8 & 4& 4& 3& 3& 4& 3& 0& 3      \\	
                            \hline
			$\!\!$umuCD$\!\!$ & 8 & 7& 5& 5& 4& 4& 5& 2& 0      \\										
						\hline					

			\end{tabular}\label{table:Perturbance}
\end{table}

\renewcommand{\baselinestretch}{1.45}


\end{document}